# Observation of electronic nematicity driven by three-dimensional charge density wave in kagome lattice $KV_3Sb_5$

Zhicheng Jiang,[†,‡,#] Haiyang Ma,[fj,§,#] Wei Xia,[fj,§,#] Zhengtai Liu,[†,‡,#] Qian Xiao,[∥] Zhonghao Liu,[†,‡] Yichen Yang,[†,‡] Jianyang Ding,[†,‡] Zhe Huang,[†,fj] Jiayu Liu,[†,‡] Yuxi Qiao,[†,‡] Jishan Liu,[†,‡] Yingying Peng,[∥] Soohyun Cho,[*,†] Yanfeng Guo,[*,fj,§] Jianpeng Liu,[*,fj,§] and Dawei Shen[*,⊥,†]

[†]*State Key Laboratory of Functional Materials for Informatics, Shanghai Institute of Microsystem and Information Technology, Chinese Academy of Sciences, Shanghai 200050, China*

[‡]*Center of Materials Science and Optoelectronics Engineering, University of Chinese Academy of Sciences, Beijing 100049, China*

[fj]*School of Physical Science and Technology, ShanghaiTech University, 201210, Shanghai, China*

[§]*ShanghaiTech Laboratory for Topological Physics, ShanghaiTech University, 201210, Shanghai, China*

[∥]*International Center for Quantum Materials, School of Physics, Peking University, 100871, Beijing, China*

[⊥]*National Synchrotron Radiation Laboratory, University of Science and Technology of China, 42 South Hezuohua Road, Hefei, Anhui 230029, China*

[#]*Equal contributions*

E-mail: shcho@mail.sim.ac.cn; guoyf@shanghaitech.edu.cn; liujp@shanghaitech.edu.cn; dwshen@ustc.edu.cn

**Abstract**

Kagome superconductors AV$_3$Sb$_5$ (A = K, Rb, Cs) provide a fertile playground for studying intriguing phenomena, including non-trivial band topology, superconductivity, giant anomalous Hall effect and charge density wave (CDW). Recently, a $C_2$ symmetric nematic phase prior to the superconducting state in AV$_3$Sb$_5$ drew enormous attention due to its potential inheritance of the symmetry of the unusual superconductivity. However, direct evidence on the rotation symmetry breaking of the electronic structure in the CDW state from the reciprocal space is still rare, and the underlying mechanism remains ambiguous. The observation shows unconventional unidirectionality, indicative of rotation symmetry breaking from six-fold to two-fold. The interlayer coupling between adjacent planes with $\pi$-phase offset in the 2×2×2 CDW phase leads to the preferred two-fold symmetric electronic structure. These rarely observed unidirectional back-folded bands in KV$_3$Sb$_5$ may provide important insights into its peculiar charge order and superconductivity.

Owing to the intrinsic geometric frustration in real space and non-trivial band topology in reciprocal spaces, compounds with kagome lattices may exhibit various novel quantum states of matter, such as kagome magnet,[1–8] flat bands,[9,10] fractional Chern insulator states [11] and superconductivity. [10,12,13] In this regard, the recent discovery of a new family of kagome superconductors AV$_3$Sb$_5$ (A=K, Rb and Cs) has naturally attracted great interest.[13–29] Subsequent researches soon dug out rich exotic states in this family of kagome superconductors, including the nontrivial Z$_2$ topological surface states, [13,20] strong-coupling superconductivity with possible triplet pairing[20,21,30,31] and giant anomalous Hall effect (AHE) without observable long-range magnetic orders.[32,33]

Besides, there exist charge density wave (CDW) transitions in all of the AV$_3$Sb$_5$ systems, which have been observed by kinks in the specific heat, then subsequently confirmed by static charge modulations in the scanning tunneling microscopy/spectroscopy (STM/S) occurring at 80 ~ 100 K.[13–19,28] Right now, the symmetries of the electronic structure in the CDW phase of AV$_3$Sb$_5$ are still under debate. Some magnetoresistance measurements and STM/S results provide evidences on the two-fold ($C_2$) rather than six-fold ($C_6$) symmetric electronic ground state in AV$_3$Sb$_5$, which was suggested to be attributed to either an intrinsic nematic order or incommensurate three-dimensional charge order in their CDW states.[19,22,23,29,34,35] Given that other ordered electronic states, e.g., the superconductivity sets in on the base of the charge order, they would logically inherit the vestige of symmetries and properties of the CDW phase. Thus, the newly discovered symmetry-broken electronic structure can shed light on some key information of these ordered states including the symmetry of superconducting order parameter. On the other hand, the latest magnetic-field dependent STM and muon spin relaxation/rotation ($\mu$SR) measurements on AV$_3$Sb$_5$ demonstrated the chiral charge order and time-reversal symmetry (TRS) breaking associated with orbital currents in their CDW states, which were proposed to account for the exotic giant AHE without observable long-range magnetic order.[15,24,25] We note that although various competing charge orders with distinct symmetries and their interplay have been under intense scrutiny through STM/S, $\mu$SR, and transport measurements, the evidence on the symmetry breaking of electronic structures of AV$_3$Sb$_5$ in the reciprocal lattices, which could provide more compelling and straightforward symmetry information of electronic states, is still lacking.[19,24,29]

In this Letter, utilizing the micron-scale spatially resolved angle-resolved photoemission spectroscopy (ARPES), we reveal an intrinsic rotation symmetry breaking of the band structure in the CDW phase of KV$_3$Sb$_5$. We have successfully observed the fingerprint of the unidirectional band folding in the CDW phase of KV$_3$Sb$_5$, which yet emerges just along one of three equivalent 2×2 in-plane CDW wavevector directions, unambiguously indicating that the rotation symmetry has been reduced from $C_6$ to $C_2$. Combining first principles calculations and high-resolution X-ray diffraction (XRD), we pinpoint that the interlayer coupling between adjacent planes with $\pi$-phase offset in the three dimensional 2×2×2 inverse star- of-David CDW phase[36–39] would lead to one unique direction, thus breaks the rotational symmetry. Our findings shed light on the subtle connections and interplay between electronic structure, nematicity, charge order in this class of complex kagome metals, which may help to further understand the mechanism of CDW and superconductivity therein.

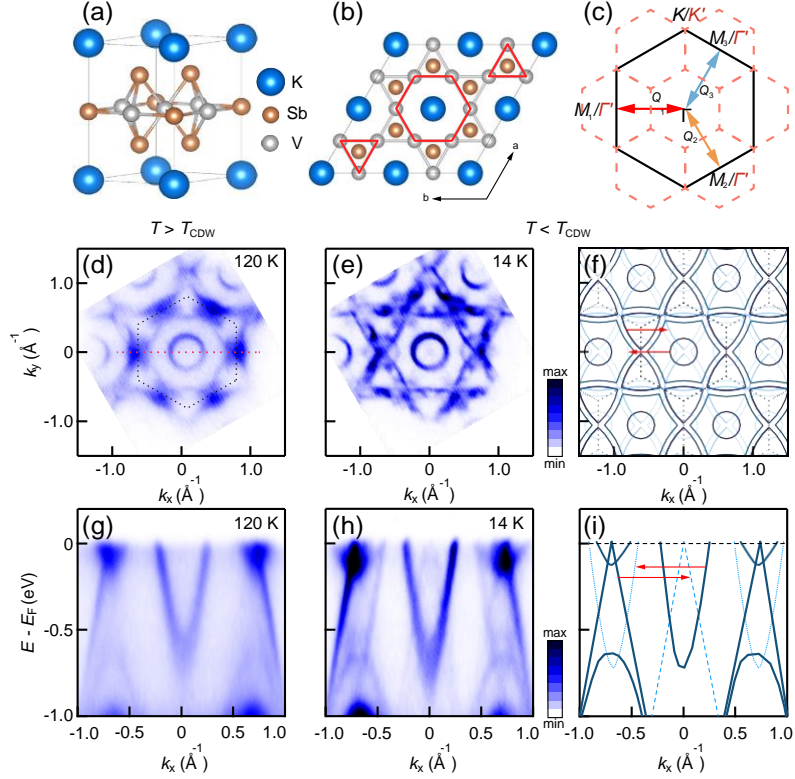

Figure 1: (a) The structural unit cell of $KV_3Sb_5$: K, V, Sb atoms are presented as blue, gray and gold balls, respectively. (b) Top view of kagome plane in $2a_0 \times 2a_0$ unit cell of $KV_3Sb_5$. Gray and Red frames represent the bonds of V atoms in normal and CDW states, respectively. (c) Illustration of pristine (dark solid line) and in-plane 2×2 (red dash line) Brillouin zones. The three colored arrows denote the scattering directions. $\Gamma$, $K$, $M$ ($\Gamma'$, $K'$) represent the high symmetry points of pristine (in-plane 2×2) Brillouin zones. (d, e) Fermi surface ARPES mapping taking at 120 K ($T > T_{CDW}$) and 14 K ($T < T_{CDW}$). (f) Sketch of band structure imitated from Figure 1e. (g, h) Band dispersion measured at 120 K and at 14 K along the high symmetry $M$-$\Gamma$-$M$ direction marked by red line in Figure 1d. (i) Sketch of pristine and folding band dispersion of Figure 1h. The solid (dash) lines denote the pristine (back-folding) bands.

$KV_3Sb_5$ crystallizes in a layered kagome lattice structure with the space group $P6/mmm$ (No. 191), as illustrated in Figure 1a. $KV_3Sb_5$ enters the CDW state at $\sim$78 K, in which V-Sb kagome lattices exhibit the in-plane 2×2 inverse star-of-David (ISD) charge order distortion (Figure 1b).[14,37] Subsequently, as shown in Figure 1c, reduced CDW Brillouin zones (BZs) form in the reciprocal space with three equivalent reciprocal vectors, by which the band folding is expected to emerge. In Figures 1d and 1e we directly compare photoemission intensity maps at the Fermi level ($E_F$) taken in the normal (120 K) and CDW phases (14 K) of KV$_3$Sb$_5$, respectively. In the CDW state (Figure 1e), the intensity map demonstrates evident band folding induced extra shadow rings around some of $M$ points and 'X'-like features around Γ, which were rarely reported in previous ARPES works on AV$_3$Sb$_5$.[40] A more detailed comparison between band dispersions along $k_y$ = 0 (the red dashed line in Figure 1d) in the normal and CDW phases further confirms these CDW induced back-folded bands (Figures 1g and 1h) and additional hole and electron pockets with relatively weak spectral weight appear around the Γ and $M$ points, respectively. Such phenomena are rather reminiscent of back-folded bands observed in transition metal chalcogenides such as 1$T$-TiSe$_2$, CeTe$_3$ and 2$H$-TaS$_2$,[41–43] which have been regarded as the smoking gun for the presence of CDW states in photoemission. From the sketches imitated from experimental data, as illustrated in Figures 1f and 1i, we can unambiguously attribute them to back-folded spectra from original bands in the normal phase by the in-plane wavevector $\boldsymbol{Q}_1$ (the red double sided arrow in Figure 1c).

It is worthwhile to point out that the rotation symmetry in KV$_3$Sb$_5$ reduces from $C_6$ to $C_2$ upon the CDW transition accompanied by the emergence of back-folded bands. Given the kagome lattice, the electronic structure of KV$_3$Sb$_5$ naturally inherits the $C_6$ rotational symmetry. However, the additional back-folded bands just appear along one ($\boldsymbol{Q}_1$) rather than the other two reciprocal-vector directions (Figures 1e-f), which demonstrates the rotation symmetry breaking of the electronic ground state in the CDW phase. To further verify such a rotation symmetry breaking in KV$_3$Sb$_5$, we directly compare the photoemission intensity plots taken along different CDW reciprocal-vector directions Γ-$M_1$ (Figure 2b) and Γ-$M_3$ (Figure 2c). Note that these two directions are equivalent in the $C_6$ preserving BZ of the normal phase, but not commensurate in the rotation-symmetry-broken $C_2$ one, as illustrated in the upper panel of Figure 2a. These spectra exhibit distinctly different band structures around $M_1$ (Γ̀) and $M_3$ (S) points. While the back-folded electron pocket from Γ can be well identified around the $M_1$ point (marked by red arrows in Figure 2b), no sign of such band folding is visible around $M_3$ as illustrated in Figure 2c. Meanwhile, the $C_6$ symmetry breaking of folding bands can also be seen along directions deviated from high symmetry lines (Figure S1) and CDW gaps (Figures S7 and S8). Figures 2d and 2e present constant energy photoemission intensity maps at 0.2 and 0.6 eV below $E_F$, respectively, which more clearly show the extra band-folding induced rings and 'X'-shape features with respect to $E_F$ mapping. As well, we found that back-folded rings are only located around $M_1$ rather than $M_2/M_3$. Furthermore, as shown in Figure 2f, one photoemission intensity plot taken along the close loop (as indicated by the orange dashed circle with a radius of 0.42 Å$^{-1}$ in Figure 2d), which should cross all back-folded bands around Γ, indicates that there exist just four instead of six hole-like bands enclosing the zone center. Together with the four pairs of back-folded-band peaks shown in the corresponding momentum distribution curve taken at 0.2 eV below $E_F$, we can ascertain the absence of $C_6$ rotation symmetry in the CDW phase of this compound. We emphasize that such unidirectionality is not a trivial consequence of the matrix element effect in photoemission spectroscopy. We have ruled out this possibility by varying the measurement geometry, after rotating the sample by 60° clockwise, these back-folded bands rotate by 60° as well accordingly and still remain unidirectional (Figure 2g and momentum distribution curves (MDCs) in Figure 2f). Moreover, we find the orientation of $C_2$ folding bands rotates by 60° on different domains of same sample, which is similar to the discovery in previous STM reports[23] and further exclude the influence of matrix element effect (see details in section 1 of Supporting Information (SI)).

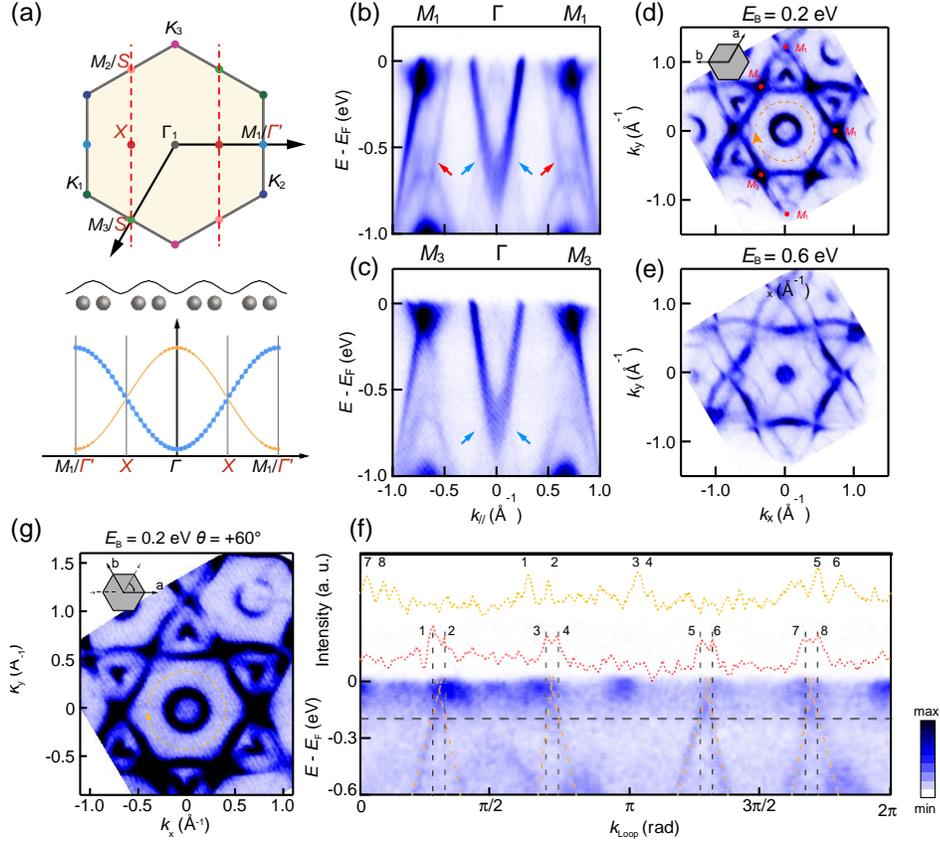

Figure 2: (a) Upper: Illustration of pristine hexagonal Brillouin zone, the colored dots denote anisotropy high symmetry points in CDW state; Lower: Illustration of CDW induced back-folded bands in pristine period. (b), (c) High symmetry cut spectras along the $M_1$-$\Gamma$-$M_1$, $M_3$-$\Gamma$-$M_3$ direction. Folding bands around $\Gamma$ and $M_1$ are denoted by blue and red arrows, respectively. (d), (e) ARPES spectras of constant energy surface mapping at binding energy of 200, 600 meV, taken with $h\nu$ = 88 eV (inset on Figure 2d top left corner illustrates the sample configuration). (f) Band dispersion cut along the circular loop marked by origin curve in Figure 2d. The corresponding momentum distribution curve (red dash line) taken at binding energy of 200 meV is appended in same subfigure. (g) Constant surface mapping of the sample rotated clockwise by $\theta$ = 60° with respect to the configuration in Figure 2d. The corresponding momentum distribution curve along the circular loop marked by yellow dash line is appended in the uppermost of Figure 2f.

Prior researches regarded the 2×2 ISD structure to be the most stable in-plane CDW reconstruction in AV$_3$Sb$_5$.[37] Later, the 2×2×2 three-dimensional CDW configuration was reported in some AV$_3$Sb$_5$ compounds.[20,35,37,44–49] To investigate impacts on the rotation symmetry of electronic structure upon CDW transitions, we compared side-by-side both the schematics of the back-folded bands induced by different stacking structures (Figures 3a-b), and detailed band structure calculations in the two distinct CDW structures (Figures 3c-d). Figure 3a shows the 2×2 CDW stacking configuration, in which the $C_6$ rotation symmetry inherited from the kagome lattice would be preserved. Consequently, the band folding along all three CDW wavevector directions should be equivalent and non-unidirectional. The calculated constant energy map at $E_B$ = 0.1 eV as well demonstrates the $C_6$ rotation symmetric nature (Figure 3c). In contrast, for the three-dimensional $\pi$-ISD CDW phase (Figure 3b), the bottom ISD sheet in the CDW supercell would be shifted along one of the three pristine lattice vectors with respect to the top on.[35] Without loss of generality, here we assume that the two kagome layers are shifted by one pristine lattice vector along the $a$ axis (see Figure 3b). As a result, the lattice potential would have a relatively strong modulation along the direction perpendicular to $a$ axis, which has a modulation length of $\sqrt{3}a$, and corresponds to one of the three reciprocal vectors $\mathbf{Q}_1 = [-2\pi/\sqrt{3}a, 0, \pi/2c]$. This would lead to an increased Fourier component of the lattice potential at the reciprocal vector $\mathbf{Q}_1$, which enhances the scatterings and band folding effects between $\Gamma(A)$ and $L(M)$ pockets along the $\mathbf{Q}_1$ direction. It is noted that the $k_z$ dispersion is weak along $\Gamma - A$ and $M - L$, the electronic structure around $\Gamma(M)$ and $A(L)$ are nearly identical, we just discuss the scattering from $\Gamma$ to $M$, as shown in Figure S11. Thus, the unidirectional crystal field with the new period would emerge upon the CDW formation, reducing the original $C_6$ to the $C_2$ rotation symmetry in the reciprocal space.[34,35,48] Note that the resulting constant energy photoemission intensity map calculation in Figure 3d shows such a rotation symmetry breaking, and we can recognize the ring-like feature only around $M_1$ and the two-fold symmetric 'X'-like spectra around $\Gamma$. Additionally, the anisotropic back-folded spectral functions along $\Gamma$-$M_1$ and $\Gamma$-$M_{2,3}$ can also be recognized by a continuum model, as shown in Figures 3e and 3f (see more details in Figures S17-S18). Furthermore, an increase in the interlayer coupling parameter would give rise to the dissimilarity of the band dispersion (see inset of Figure 4c). These findings provide strong evidence on a three-dimensional CDW distortion driving $C_2$ electronic state in $KV_3Sb_5$. Therefore, we expect that 2×2×2 CDW transition would enhance the electronic scattering to a preferred 1D direction, thus generating the electronic nematic phase in the kagome lattice. We as well expect that the $C_2$ electronic nematicity in the CDW phase might determine the charaterisitics of superconductivity, such as the superconducting gap symmetry, at lower temperatures.

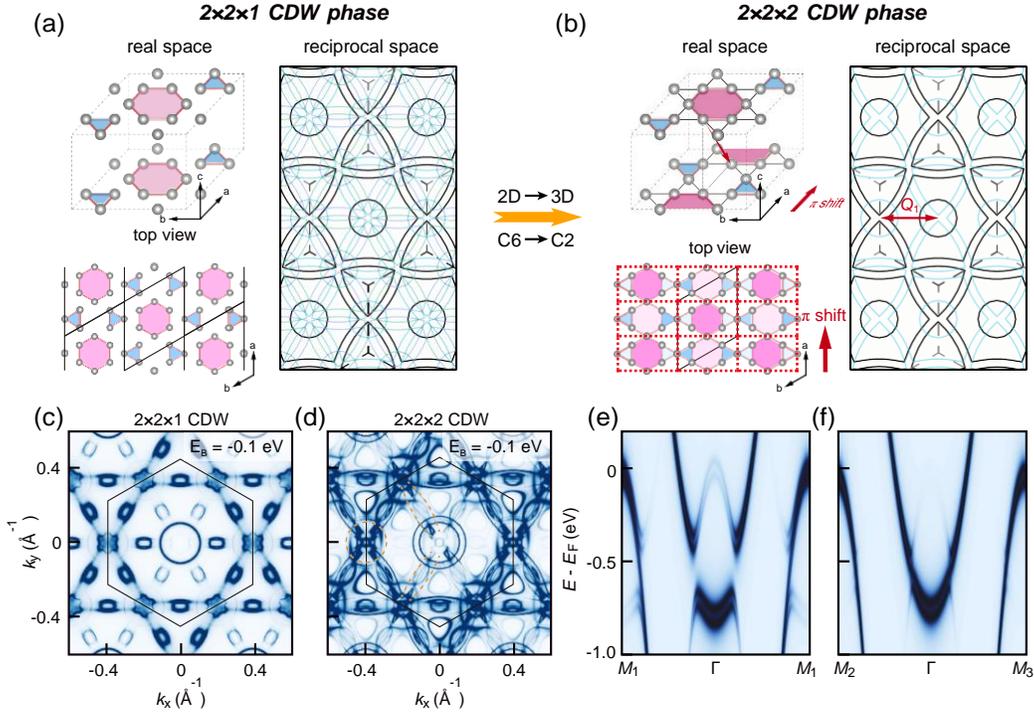

Figure 3: (a, b) Left: Side and top view of V net bilayer of inverted Star-of-David (ISD) (a) 2×2×1 ((b) 2×2×2, red arrows denote the interlayer shift direction) CDW phase (dark diamond frames denote the 2×2×1 super cells); Right: Distribution of pristine and (a) 2×2×1 ((b) 2×2×2) CDW folding bands in 2D projected reciprocal space. (c, d) DFT Calculated intensity map taken at $E_B = 0.1$ eV from ISD 2×2×1 (2×2×2) CDW structure(Dark hexagon represent the pristine Brillouin zone). (e) and (f) Back-folded spectral function calculated from a continuum model (see SI section 2.2).

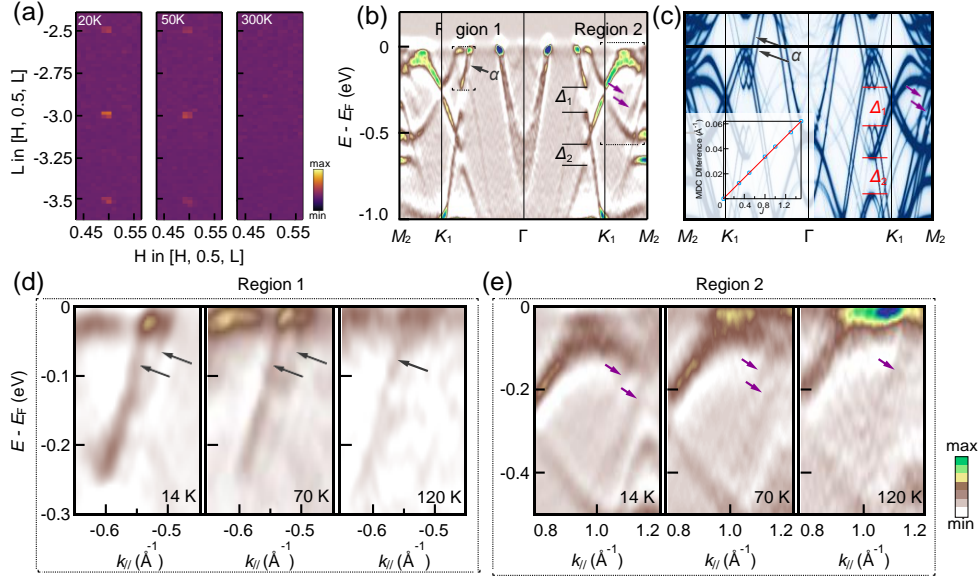

Figure 4: (a) Narrow [H, L] maps of reciprocal space of the CDW diffractions in warming process. (b) Second derivative spectra along $M_2$-$K_1$-$\Gamma$-$K_1$-$M_2$ direction taken at 14 K. (c) Calculated back-folded band dispersion for the 2×2×2 CDW phase along $M_2$-$K_1$-$\Gamma$ direction, $\Delta_1$ and $\Delta_2$ mark the folding gaps. The gray arrows denote the splitting $\alpha$ bands and purple arrows denote the extra band dispersion near $M$ point in 2×2×2 CDW phase. The inset in Figure 4c shows the dependence of the CDW-induced band splitting (marked by the double black arrows) on the interlayer coupling scaling parameter $J$, with $J = 1$ corresponding to the physical interlayer coupling strength. (d), (e) Enlarged second-derivation images from the region 1 and region 2 in Figure 4b. The parallel three spectra in Figure 4d and 4e are taken at 14K, 70 K and 120 K, respectively.

We then performed a detailed temperature dependent measurement to confirm such a three-dimensional CDW transition. From single crystal XRD in Figure 4a, we can distinguish CDW induced peaks at both integer and half-integer $L$ points below $T_{CDW}$, which unambiguously indicates the existence of such a 2 × 2 × 2 CDW transition in $KV_3Sb_5$. The second derivative spectra in Figure 4b shows the band splitting in regions 1 and 2, which are characteristic of the 2×2×2 three-dimensional CDW phase, and in marked agreement with the calculation (Figure 4c and Figure S16). We note that the magnitude of splitting for bands in both regions 1 and 2 should be proportional to the interlayer coupling of V-Kagome layer, and temperature dependent evolution of the band splitting can identify the three-dimensional CDW onset temperature (Figures 4d-e and more details about photon energy and temperature dependent spectra can be found in Figures S9-S14). We discovered that KV$_3$Sb$_5$ enters the CDW phase with a 2×2×2 supercell just below $T_{CDW}$. This finding coincides with the $C_6$ rotational symmetry breaking (or nematic phase), which should have a profound effect on the superconductivity occurring at lower temperatures. Figure S19 presented temperature dependent measurements of Hall effects in KV$_3$Sb$_5$. The feature of anomalous Hall effect emerges at low temperatures ⪅ 40 K. The electron concentrations ($n_e$) and mobilities ($\mu_e$) extracted from the ordinary Hall effect also show kinks at ~ 50 K. These results imply that there may exist additional phase transition at lower temperatures, which further break the time-reversal symmetry within the nematic CDW phase, which proposed to exhibit counter-propagating current loops as suggested by $\mu$SR measurements[24,25] and theoretical calculation.[50] Thus, the low-temperature photoemission spectra discussed above directly demonstrate the electronic structure of a peculiar quantum state of matter with co-existing TRS breaking and $C_6$ breaking nematic orders.

We have directly observed the rotation symmetry breaking of the electronic structure upon the CDW transition of KV$_3$Sb$_5$ in its reciprocal space. The symmetry broken phase indicates the existence of electronic nematic ordering which onsets at $T_{CDW}$. It is driven by the three-dimensional CDW transition with $\pi$ phase shift rather than in-plane anisotropic instability of electrons below $T_{CDW}$. This finding is consistent with several recent reports.[20,23,44–46] However, we note that the track of band folding in AV$_3$Sb$_5$ is absent in most previous ARPES reports.[13,27,28,45,51–56] We speculate that the absence of such characters might result from the inhomogeneous cleavage plane, over which domains with preferred directions in the CDW state are spread randomly.[26] In this regard, the micron-scale spatially resolved ARPES is crucial in probing the unidirectionality of the nematic CDW phase from the single domain in $AV_3Sb_5$.

On the other hand, the CDW state in $KV_3Sb_5$ is much more homogeneous. Our XRD results have revealed a single 2×2×2 phase transition in $KV_3Sb_5$. However, in $CsV_3Sb_5$, it undergoes an additional quadrupling structural modulation along the *c*-axis,[46] which would compete with the 2×2×2 CDW phase formed at 94 K. Moreover, the temperature evolution of anisotropic 2×2 CDW peaks and the V splitting in nuclear magnetic resonance measurements both indicate the additional nematic transition occurring at ~40 K. To date, it is still unknown whether this nematic order emerging at lower temperature in $CsV_3Sb_5$ would be intertwined with or compete against the nematic charge order induced by the three-dimensional CDW. Furthermore, the additional $4a_0$ stripe order has been observed in $AV_3Sb_5$ except for $KV_3Sb_5$,[16–19,57,58] and the multiple competing charge orders in $(Cs,Rb)V_3Sb_5$ would make the nematic electronic states hard to be observed in ARPES.

In conclusion, we have characterized the electronic structure with unique three-dimensional CDW phase in $KV_3Sb_5$. Though a comprehensive analysis of the $C_2$ symmetric band dispersion and comparision with the geometrical configuration of the reciprocal space, we can draw a direct connection between the rotational symmetry broken electronic structure in reciprocal space and the $\pi$-ISD three-dimensional CDW phase in real space. Meanwhile, the temperature dependence of the XRD and the band splitting in ARPES spectra further indicate the single 2×2×2 CDW transition in $KV_3Sb_5$. The anomalous Hall effect induced by weak magnetic fields emerges at lower temperatures $\lessapprox$ 40 K, which may imply additional TRS breaking phase transition within the nematic CDW phase. Thus, our low-temperature ARPES spectra unveils the electronic structure of a peculiar nematic CDW phase with broken TRS. These findings shed light on the complex and puzzling properties of the nematic phase, charge order, and time-reversal symmetry breaking in $AV_3Sb_5$ systems, which would provide useful guidelines for understanding the mechanism of superconductivity in these V-based kagome superconductors.

Angle-resolved photoemission (ARPES) measurements were performed at the BL03U

beamline of Shanghai Synchrotron Radiation Facility (SSRF), [59] the light is set to linear horizontal polarization (parallel to the ground), and the beam spot at the sample position is round with size of 15 ≈ 15 $\mu m^2$. The ARPES end station is equipped with a Scienta-Omicron DA30 analyzer with vertical slit. In the experiment process, the slit was set to 300 $\mu m$ and all data is taken with 88eV photon in main text. $KV_3Sb_5$ single crystal sample is glued to the anaerobic copper sample holder with silver adhesive, and then cleaved *in situ* at 10 K with a base pressure of better than 8 × $10^{-11}$ Torr. The samples were heated up to a maximum of 150 K during measurements. The total energy resolution was set to better than 20 meV and angular resolution was set to be 0.2°, respectively.

First-principles calculations based on density functional theory (DFT) are performed using the Vienna ab initio simulation package (VASP) which adopts the projector-augmented wave method.[60] The energy cutoff is set at 400 eV and exchange-correlation functional of the Perdew-Burke-Ernzerhof (PBE) type[61] is used for both the structural relaxations and electronic structures calculations. The convergence criteria for the total energy and forces are set to $10^{-6}$ eV and 0.001 eV/Å, respectively. The BZ is sampled by a 9×9×7 **k** mesh for the pristine structure, and a 6×6×4 **k** mesh for the CDW structures. Unfolded spectral functions are calculated based on the Wannier tight-binding models which are obtained through the *wannier90* code.[62] Specifically, the unfolded spectral function at energy $\omega$ and wavevector **k** is expressed as $A(\omega, \mathbf{k}) = -(1/\pi) \operatorname{Im} \sum_n |W_{n\mathbf{k}}|^2 / (\omega - E_{n\mathbf{k}} + i\eta)$, where $E_{n\mathbf{k}}$ is the band energy, $\eta$ = 0.01 eV is a small smearing factor, and $W_{n\mathbf{k}}$ is the unfolded spectral weight which are obtained by projecting the Bloch functions in the CDW phase onto those of a 1×1×2 structure, which has a $\pi$ shift between the two adjacent kagome layers.[63,64]

We performed single crystal x-ray diffraction (XRD) measurements by using a custom-designed x-ray instrument. It was equipped with a low divergence channel cut monochromator and a Xenocs Genix3D Cu $K_{\alpha 1}$ (8.048 keV) x-ray source, which provided a beam spot size of 0.5 mm × 1 mm at sample position with 4.6 × $10^6$ photons/sec. The measured samples were mounted on a Huber 4-circle diffractometer. A highly sensitive PILATUS3 R 1M solid state pixel array detector with 980 × 1042 pixels was used to collect the scattering signals. Each pixel size was 172 $\mu$m × 172 $\mu$m. The 3D mapping of momentum space was obtained by taking images with an increment of 0.1° in sample-rotations.

## Acknowledgement

The authors thank Prof. Zhenyu Wang for valuable discussions. The authors acknowledge the support by the National Natural Science Foundation of China (Grants No. U2032208, 92065201, 11974029, 12222413 and 12174257), the Shanghai Science and Technology Innovation Action Plan (Grant No. 21JC1402000), the Natural Science Foundation of Shanghai (Grant No. 22ZR1473300 and 23ZR1482200), the National Key R & D program of China (Grant No. 2020YFA0309601, 2022YFB3608000) and the Double First-Class Initiative Fund of ShanghaiTech University. Part of this research used Beamline 03U of the Shanghai Synchrotron Radiation Facility, which is supported by ME$^2$ project under Contract No. 11227902 from National Natural Science Foundation of China.

## Supporting Information Available

Additional experimental data, including the detailed experimental methods, analysis of matrix element effect of photoemission, spatial scannning maps, temperature dependent and photon energy dependent ARPES measurements, calculation details and Hall measurement results.

# Supporting Information for

# "Observation of electronic nematicity driven by three-dimensional charge density wave in kagom lattice KV$_3$Sb$_5$"

Zhicheng Jiang,[†,‡,#] Haiyang Ma,[fj,§,#] Wei Xia,[fj,§,#] Zhengtai Liu,[†,‡,#] Qian Xiao,[∥] Zhonghao Liu,[†,‡] Yichen Yang,[†,‡] Jianyang Ding,[†,‡] Zhe Huang,[†,fj] Jiayu Liu,[†,‡] Yuxi Qiao,[†,‡] Jishan Liu,[†,‡] Yingying Peng,[∥] Soohyun Cho,[*,†] Yanfeng Guo,[*,fj,§] Jianpeng Liu,[*,fj,§] and Dawei Shen[*,⊥,‡]

[†]State Key Laboratory of Functional Materials for Informatics, Shanghai Institute of Microsystem and Information Technology, Chinese Academy of Sciences, Shanghai 200050, China

[‡]Center of Materials Science and Optoelectronics Engineering, University of Chinese Academy of Sciences, Beijing 100049, China

[fj]School of Physical Science and Technology, ShanghaiTech University, 201210, Shanghai, China

[§]ShanghaiTech Laboratory for Topological Physics, ShanghaiTech University, 201210, Shanghai, China

[∥]International Center for Quantum Materials, School of Physics, Peking University, 100871, Beijing, China

[⊥]National Synchrotron Radiation Laboratory, University of Science and Technology of China, 42 South Hezuohua Road, Hefei, Anhui 230029, China

[#]Equal contributions

E-mail:dwshen@ustc.edu.cn; liujp@shanghaitech.edu.cn; guoyf@shanghaitech.edu.cn; shcho@mail.sim.ac.cn

# 1. ARPES Measurement results

## 1.1 Anisotropic band structure

Except the anisotropioc band stucture along $\Gamma$-$M_1$ and $\Gamma$-$M_{2,3}$, moreover, such a distinction in the CDW introduced band folding also exists between photoemission spectra collected along the $K_1$-$K_2$ and $K_1$-$K_3$ cuts, which are parallel to $\Gamma$-$M_1$ and $\Gamma$-$M_3$, respectively (Figure S1). Figure S1b presents the dispersion along asymmetrical direction under $C_2$ symmetry, we can distinguish the shape of folding bands near $K_3$ is hole-like pocket, but near $K_1$ is a electron-like pocket. However, from the high symmetric direction cuts $K_1$-$K_2$, the "X"-shape crossing at the center of $K_1$-$K_2$ folding from $K$ point can be clearly seen, and all the band dispersion are centrosymmetric, as shown in Figure S1c. These results also consist well with the calculation in Figure S16e and f.

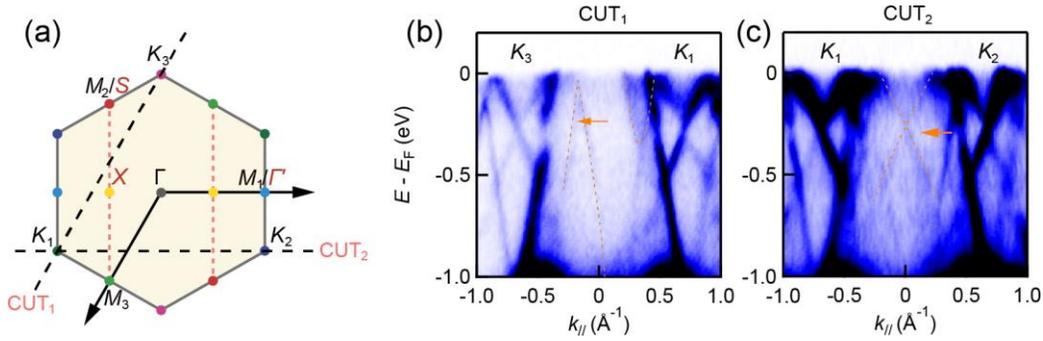

Figure S1: **Experimental band dispersion along asymmetric and symmetric *K-K'* direction.** (a) Distribution of high symmetry momentum in rotation symmetry broken Brillouin zone of pristine unit cell (marked by dark characters) and repeat rectangle period with unfold band (marked by red characters). (b), (c) Experimental band dispersion cut along asymmetric $K_3$-$K_2$ (CUT$_1$) direction and symmetric $K_2$-$K_1$ (CUT$_2$) direction.

## 1.2 Matrix element effect of photoemission

The photoemission spectroscopy intensity can be described by: $I(\mathbf{k}, E) = I_0(\mathbf{k}, v, \mathbf{A}) A(\mathbf{k}, E) f(E, T)$, in which $I_0(\mathbf{k}, v, \mathbf{A}) \propto \sum_{f,i} |M_{f,i}^k|^2$ and photoemission matrix element $M_{f,i}^k = \langle \phi_f^k | \mathbf{A} \cdot \mathbf{p} | \phi_i^k \rangle$. The $\phi_i^k$ term describes the initial state related to orbit geometry, the $\phi_f^k$ term describes the

final state identically equal to +1, $\mathbf{A} \cdot \mathbf{p}$ describes the the polarization of incident light, in our experiment configuration is mainly contributed by in-plane $\mathbf{A}_{py} \cdot \mathbf{p}_y$ and out-of-plane $\mathbf{A}_{pz} \cdot \mathbf{p}_z$. If both $Q_{1,2,3}$ exist, under the protection of both $M_x$ and $M_y$ mirror planes, all the folding bands around the 1st BZ center $\Gamma_1$ can be seen, forming a '∗'-like bands at $\Gamma_1$, while at 2nd BZ center $\Gamma_2$ only one mirror plane preserved, resulting in the 'X'-like shape back-folding bands. However, the Fermi surface of the back-folded band at the both BZs is 'X'-like shape that enables the $C_6$ rotational symmetry breaking as we mentioned in the main text. The observation of the back-folded band attributable to the ARPES spectra at the both BZ is not related to the matrix element effect about the ARPES measurements.

Additionally, we can still unambiguously exclude the influence of matrix element effects through changing the measurement geometry. As shown in Figure S2a, in the current experimental setup, the polarization direction of linear horizontal light is orthogonal to the analyzer slit. In this way, if orbital selective matrix element effects induced extrinsic symmetry modulation of ARPES data existed, the extrinsic symmetry reduction in ARPES data should be along or perpendicular to the slit direction, as illustrated in Figure S2e1. After the sample was rotated by 60°, the folded bands should still be horizontal/perpendicular to the slit direction. However, in our measurements, we found that the $C_2$ folding bands no more parallel/perpendicular to silt direction [Figure S2f2]. Thus, the symmetry driven by photoemission matrix element effect can be excluded without polarization-dependent experiments.

Furthermore, we also performed systematic ARPES measurements on more $KV_3Sb_5$ samples, and we have discovered that the $C_2$ band folding in $KV_3Sb_5$ is actually along different direction depending on different samples [Figure S2g1, -60° to the slit direction] and as well different domains of one sample [Figure S2g2 and g3, 60° and 0° to the slit direction, respectively]. If such a $C_6$ to $C_2$ symmetry breaking was induced by matrix element effects, the discovered folded bands in the momentum space should be independent on samples or domains, which is obviously inconsistent with our result. Our findings thus suggest that the anisotropic folded bands discovered in $KV_3Sb_5$ should be attributed to the intrinsic $C_6$ to $C_2$ symmetry breaking of electronic structure rather than extrinsic matrix element effects.

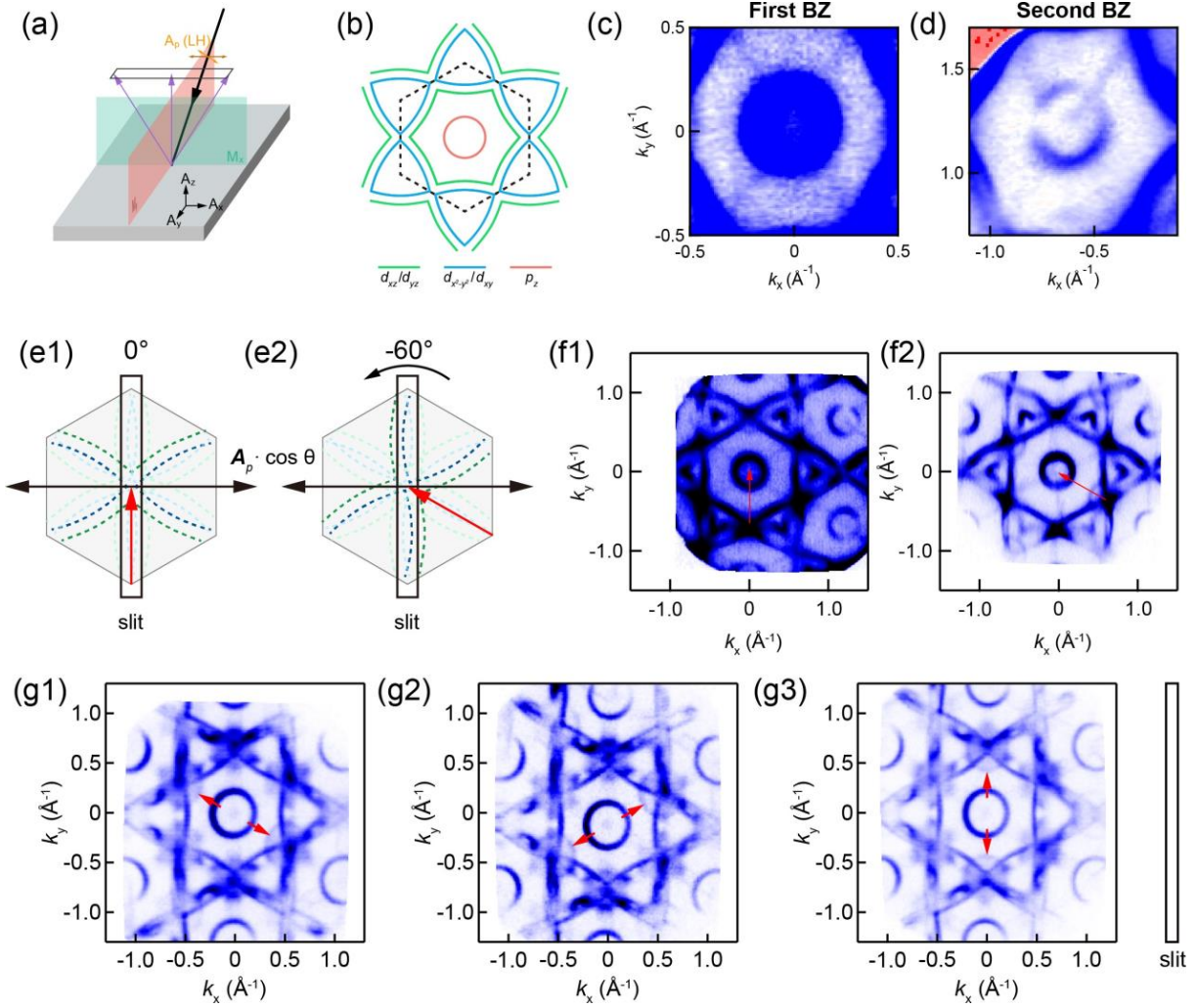

Figure S2: **Experimental geometry, orbital character and ARPES spectra in the first and second BZ.** (a) Experimental geometry for ARPES measurements where the green region is the experimental mirror plane. (b) The schematic picture of the Fermi surface of $AV_3Sb_5$ kagome lattice. The color indicate the orbital character of the atoms; the red line is the $p_z$ orbital of Sb atoms, the green line is the $d_{xz}$ and $d_{yz}$ orbital of V atoms, and the blue line is the $d_{x^2-y^2}$ and $d_{xy}$ orbital of V atoms. The band dispersion color-coded with the green line, which consist with the out-of-plane orbital character of V atoms, is represented on the "$M$-centered $\alpha$ band" in the main text and the ribbon-like shape at the Fermi surface. (c) and (d) show the Fermi surface obtained from the first and second BZ, respectively. (e1-e2) Geometry of light polariton (double-headed arrow), analyzer slit (solid rectangular frame) and folded d orbits in $KV_3Sb_5$ at (f1) 0° ($\pi$-shift direction⊥ slit direction) and (f2) -60°; (g1) Fermi surface ARPES mapping taken at sample1; (g2-g3) Fermi surface ARPES mapping taken at two different domains of sample2. The corresponding slit direction of (g-i) are marked on the right side.

## 1.3 Spatial scanning of photoemission

In our experimental configuration, beam size is about 15 × 15 $\mu$m, so we can scan the sample surface to find domains. We integrated the intensity plot with a ~1 eV energy window near the Fermi surface and show the spatial maps in Figure S3b, the sample surface is not homogeneous, each domain is ~100 $\mu$m in size. In our experiment, the observed folding bands can only appear on a tiny domain.

We then scanned the typical KV$_3$Sb$_5$ cleavage surface on a 1200 × 1200 $\mu$m² area by 40 $\mu$m/step (totally 31 × 31 sampling points), as illustrated in Figure S4a. After careful searching, we could find four typical distinct domains on the cleavage plane: spectra taken on domains 1 and 2 show clear band folding but with different orientations [Figures S4c and d], while spectra taken on domains 3 and 4 do not show any sign of folding bands [Figures S4e-f]. We then checked the corresponding $K$ 3$p$ core-levels on these four kinds of domains to investigate the difference [Figure S4b]. We noticed that right $K$ 3$p$ peaks (originating from exposed $K$ atoms) taken from all four domains are located at almost the same binding energy, but those of left $K$ 3$p$'s taken from different domains show clear difference: on domains with folding bands (domains 1 and 2), $K$ 3$p$ core-level peaks are located at slightly deeper ( 10 meV) in binding energy than those taken from domains 3 and 4 (without band folding features). Given that the cleavage surface of KV$_3$Sb$_5$ constitutes of K$^+$ and (V$_3$Sb$_5$)$^-$ domains, it has been reported that the Sb-termination (V$_3$Sb$_5$)$^-$ is slightly hole-doped and the CDW therein is suppressed, while the CDW folding bands are expected to be observed on the $K$-termination. [1] Thus, on the Sb-termination, the $K$ 3$p$ core-level peak would slightly shift towards $E_F$ [green and blue lines in Figure S4b]. In this way, we can conclude that domains 1, 2 with $C_2$ folding bands belong to the K-termination, while domains 3, 4 without clear $C_2$ band folding should be related to Sb termination.

This spatially inhomogeneous surface of KV$_3$Sb$_5$ is unlikely to be introduced by gluing a sample to the holder. For the typical layered structured compound KV$_3$Sb$_5$, if the strain of silver epoxy (the glue we usually used for the sample preparation) indeed existed, it would be expected to just affect the adjacent bottom layers close to the top of sample holder. Therefore, the unidirectional strain effect would be rather negligible to top layers of the cleaved sample, from which the photoemission signal is sent out. Consequently, the symmetry breaking on photoemission spectroscopy origins from intrinsic CDW transition rather than the uniaxial strains.

We also present the reproducibility of the $C_2$ symmetric band structure for different surfaces and different cleavages in Figure S6. We can unambiguously observe the $C_2$ symmetric folding band structure on different samples which were cleaved on different dates [Figures S6a, c and e]. Meanwhile, for the comparison purpose, we as well show some distinct band structure (without the $C_2$ symmetry breaking) [Figure S6b, d and f], which were taken on different domains of the very samples. Therefore, our findings can unambiguously pin down the existence of $C_2$ symmetry breaking of electronic structure due to the three-dimensional CDW transition.

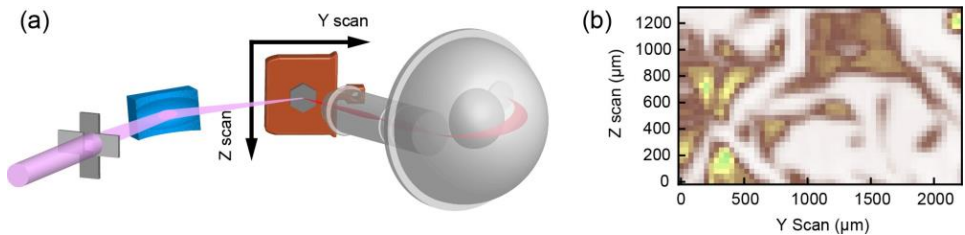

Figure S3: **Illustration of $\mu$ARPES scanning on KV$_3$Sb$_5$.** (a) Schematic of the spatially resolved ARPES in the BL03U of SSRF. (b) ARPES intensity map over the sample surface. The ARPES intensity is integrated using an energy window of ~1 eV centered at 0.5 eV below $E_F$. The surface of KV$_3$Sb$_5$ is inhomogeneous.

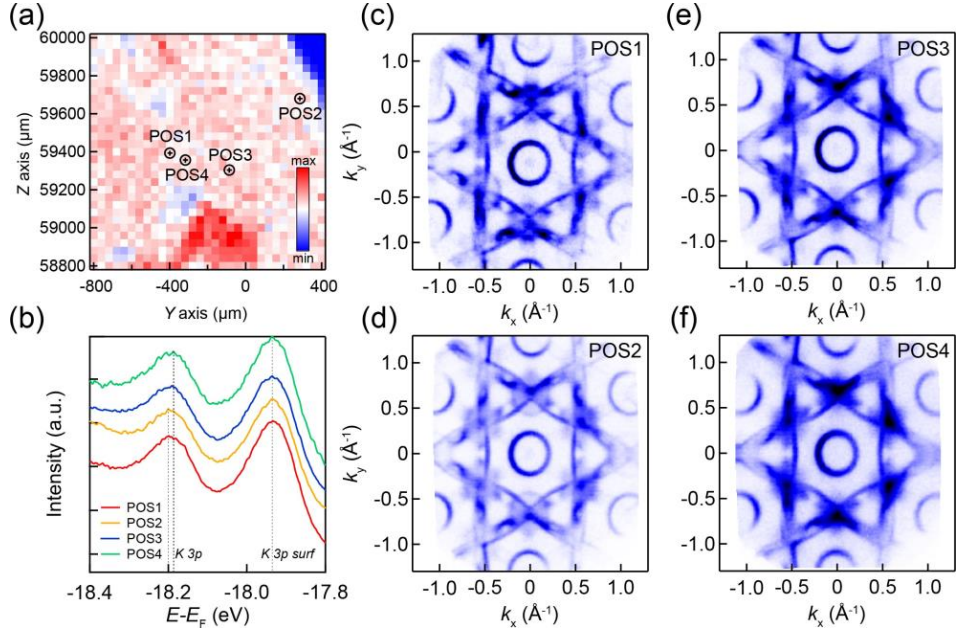

Figure S4: (a) Spatial scanning mapping of the $E_B$ position of $K\ 3p$. (b) EDCs at $K$-$3p$ core level region on different domains. (c-f) Fermi surface ARPES mapping taken at domain 1-4, respectively.

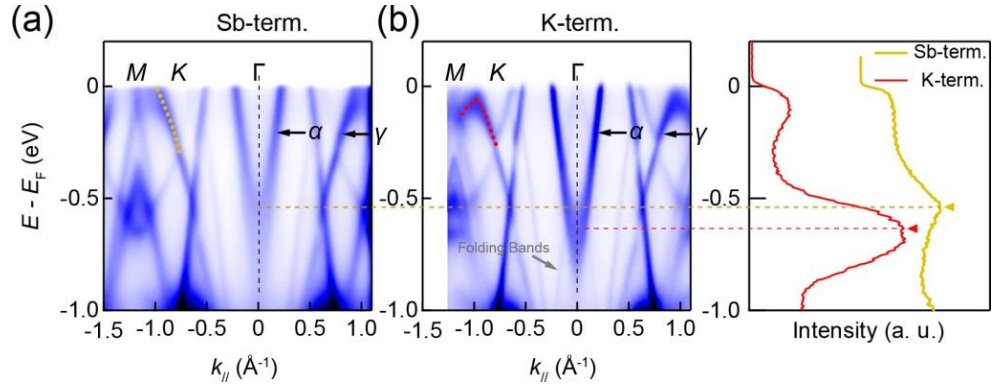

Figure S5: (a) Band dispersion cut along $M$-$K$-$\Gamma$-$K$-$M$ direction on domain without folding bands; (b) Band dispersion cut along $M$-$K$-$\Gamma$-$K$-$M$ direction on domain with folding bands; (c-e) Energy distribution curves cut along dark dash line illustrated in (a) and (b), the gold curve is extracted from (a) and red curve is extracted from (b).

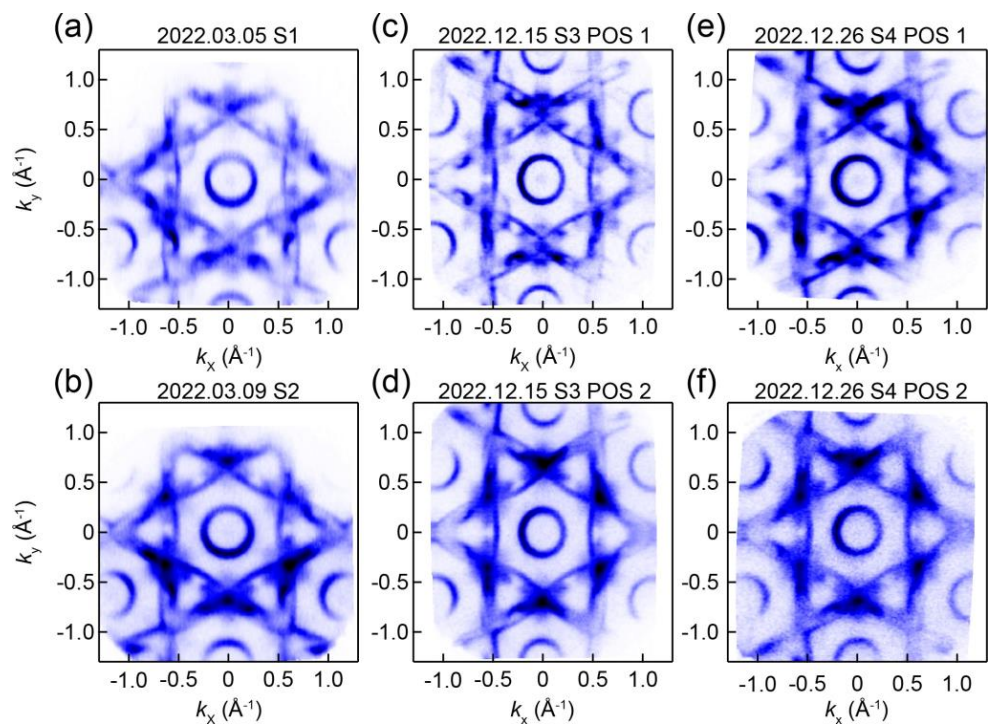

Figure S6: Fermi surface ARPES mapping taken on (a) sample 1; (b) sample 2; (c) sample 3 domain 1; (d) sample 3 domain 2; (e) sample 4 domain1; (f) sample 4 domain2.

## 1.4 CDW gaps

In this part, we discuss the CDW gap along different directions. First, we discuss the CDW gap open at saddle point near M point at $E_F$. As shown in Figure S7b-d, at the position 0.2 Å$^{-1}$ away from high-symmetric $M$ points along $M-K$ direction, there exist a hump-like band structure, where a CDW gap opens. To determine the gap size at Fermi level, we symmetrized the spectra with respect to $E_F$. We compare the gap near three different $M$ points ($M_{1,2,3}$), however, limited by the energy resolution and tiny Fermi level drifting, we found these CDW gap size did not show the apparent deviation from the $C_6$ symmetry. We speculate the CDW gap opened with intrinsic band renormalization accompanied by the CDW phase transition is isotropic, and they are dominated by in-plane 2×2 CDW phase transition.

Additionally, we indeed found those CDW gaps $\Delta_1$ induced by the anisotropic folding bands show clear anisotropic feature. The $C_2$ symmetric band features exist not only near the Fermi level, but also close to the CDW gap $\Delta_1$. Here, we extracted those ARPES intensity cuts along three nonequivalent $M$-$K$-$\Gamma$-$K$-$M$ directions taken on the domain with $C_2$ symmetric folding bands, denoted by cuts 1, 2, and 3 in Figure S8c, respectively. We note that the gap $\Delta_1$ along the cut 2 [Figure S8e] is relatively larger than those taken along the cut 1 [Figure S8d] and cut 3 [Figure S8f]. Therefore, the band folding induced CDW gap indeed deviates from the $C_6$ symmetry.

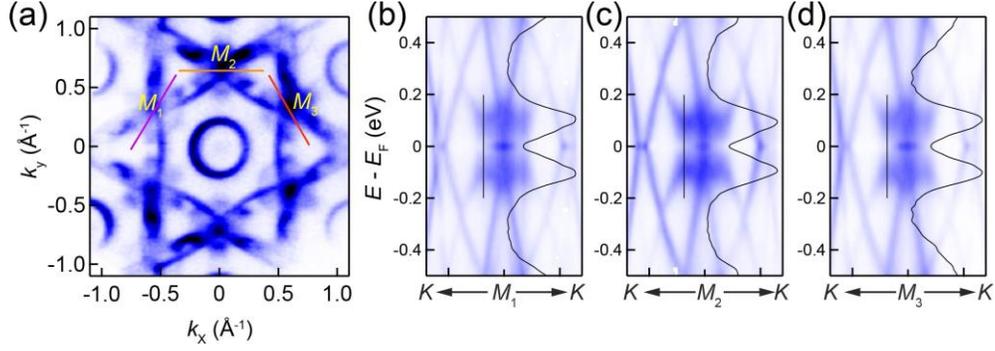

figure S7: (a) Constant energy counters of ARPES mapping taken at Fermi surface. (b-d) Band dispersion cut along $K - M - K$ direction at three different $M$ point ($M_{1,2,3}$ marked in (a)). The data were symmetrized with respect to $E_F$. The energy distribution curves across the dark straight line are appended in corresponding sub-figures.

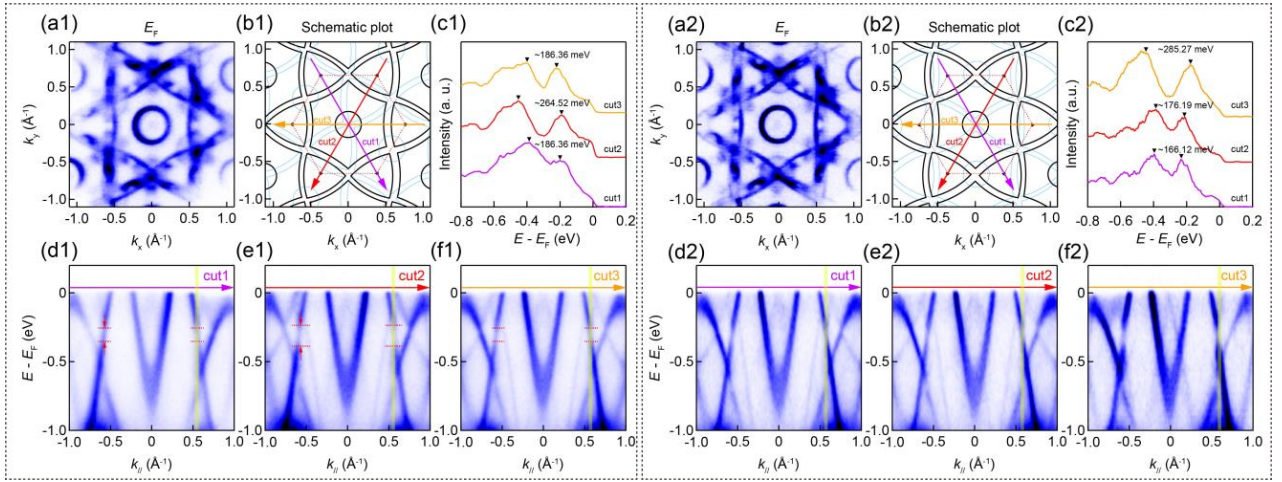

Figure S8: (a1) Constant energy counter of ARPES mapping taken at Fermi surface $E_F$; (b1) Sketch of band structure and cut directions in (d-f); (c1) The energy distribution curves (EDCs) across the CDW gap $\Delta_1$ among cut-1, 2, 3, and marked by violet, red and yellow curves, respectively; (d1-f1) Band dispersion cut along different $K$ Γ $K$ directions marked by colored arrows in (b1), the yellow shaded areas mark the integrated EDC zones of (c1). (a2) Constant energy counter of ARPES mapping taken at Fermi surface EF with C2 folding bands 60-degree respect to them in (a1); (b2) Sketch of band structure and cut directions in (d2-f2); (c2) The energy distribution curves (EDCs) across the CDW gap $\Delta_1$ among cut-1, 2, 3, and marked by violet, red and yellow curves, respectively; (d2-f2) Band dispersion cut along different $K$–Γ $K$-directions marked by colored arrows in (b2), the yellow shaded areas mark the integrated EDC zones of (c2).

## 1.5 Photon energy dependent spectra

To clarify the three-dimensional band structure of $KV_3Sb_5$, we first carried out the photon-energy-dependent $k_x$-$k_z$ map in high-symmetrized Γ-$M$-$L$-$A$ plane acquired at 1.0 eV below Fermi level [Figure S9f], we estimate the inner potential $V_0$ of $KV_3Sb_5$ is approximately 8.0 eV, which is close to the 7.3 eV of $CsV_3Sb_5$[2] and 8.2 eV of $RbV_3Sb_5$.[3]

The photon energy dependent spectra also confirm the band splitting mentioned in main text is relate to the 3D CDW. The $k_z$-$k_x$ map is acquired by varying photon energy and determine the 94 (88) eV correspond to $k_z = π$ (0) plane of 2×2×2 BZ, as shown in Figure S9f. Figure S9a and b directly compare the band dispersion from $k_z = π$ and $k_z = 0$ plane in region 1. We extract the momentum distribution curves (MDCs) from the -50 meV below Fermi level and present in Figure S9c, the band splitting is clearly on $k_z = 0$ plane but vanishes at $k_z = π$ plane. These results consist well the 3D CDW bands calculations, as shown in Figure S9d and e. This further confirmed the existence of 3D CDW phase.

Given that the CDW nesting origins from the VHSs around neighboring M points, we then measured the out-of-plane band structure in $M$ –$L$–$L'$ –$M'$ plane, marked by yellow frames as illustrated in Figure S10a. Figure S10b shows the FS topography on the yellow frame, and it is evident that the contours near the $M$ –$L$ ($M'$–$L$) line show a ×2 extra periodic variation along $k_z$ direction compared to the projected BZ, as highlighted by the dashed yellow guide line in Figure S10b, indicating the extra band folding along $k_z$ introduced by the 2×2×2 CDW transition. Meanwhile, there exist high density of states at both $M/M'$ and $L/L'$ points, which significantly increases the instability of electronic susceptibility at wavevectors connecting $M$ ($L$) and $M'(L)$. Figure 10c shows the calculated phonon spectrum of $KV_3Sb_5$, which exhibits imaginary frequencies at both $M$ and $L$, suggesting that the breathing phonon of V atoms in the kagome lattice can also result in the out-of-plane electronic instability. The strong charge fluctuations due to the Fermi-surface nesting are an important driving force for the formation of the CDW state in $KV_3Sb_5$.

We note that some other theoretical and experimental works as well concluded that the 3D electronic instability might play an essential role in driving the CDW transition,[4] and the Raman spectra have proven that the 2×2×2 superstructure in $AV_3Sb_5$ fail to induce longitudinal-acoustic and transverse-acoustic.[5] Based on our current photon-energy dependent ARPES kz dispersion, we speculate that the 3D CDW phase is driven by the Fermi surface nesting of saddle points located at different $k_z$, which are further assisted by an unstable $L$ phonon mode.

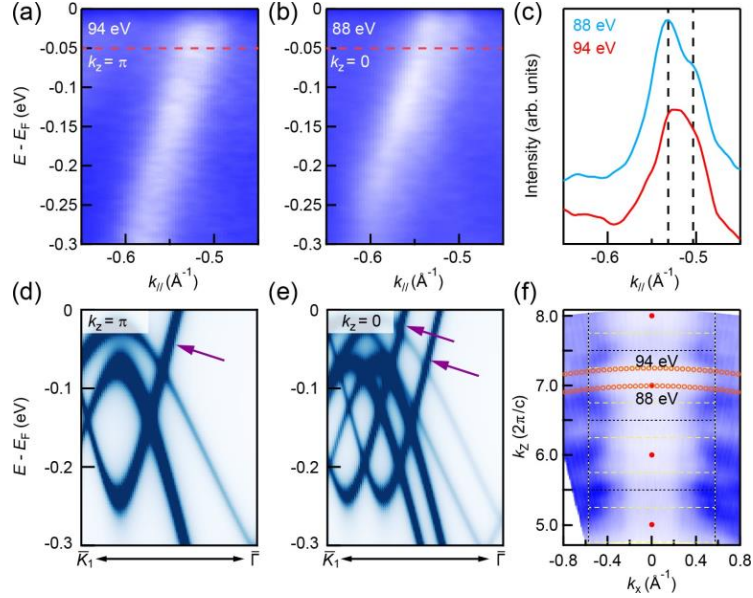

Figure S9: **Photon energy dependent evolution of band splitting.** (a), (b) Enlarged Band dispersion of region 1 in Figure 4b taken at 94 and 88 eV, which correspond to $k_z = \pi$ and 0 plane of 2×2×2 BZ respectively. (c) MDCs taken along the red dash line in (a). (d), (e) Calculated band structure of the same region in (a) and (b), purple arrows denote the splitting bands. (f) Experimental constant energy surface at $E_B$ = 1.0 eV perpendicular to the sample surface ($k_x$ vs. $k_z$ plane, $k_x$ represent the $K$-$\Gamma$-$K$ direction), showing the periodic dispersion. The dark (yellow) dash lines denote the projected pristine (2-fold reconstructed) BZ along c axis.

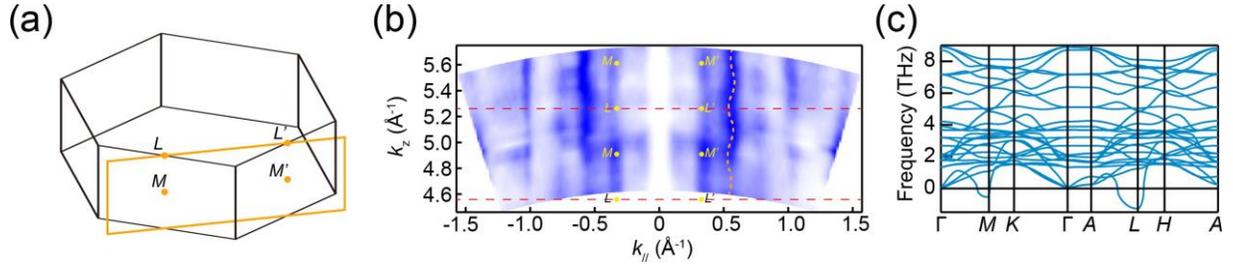

Figure S10: (a) $MLL'LM'$ plane (yellow frame) of the BZ going through the nesting region. (b) Experimental out-of-plane FS cuts in the $MLL'LM'$ plane. 3D warping of the FS counter is indicated by dashed yellow curves and Brillouin zone boundaries are indicated by red dashed lines. (c) The DFT calculated phonon spectrum of $KV_3Sb_5$.

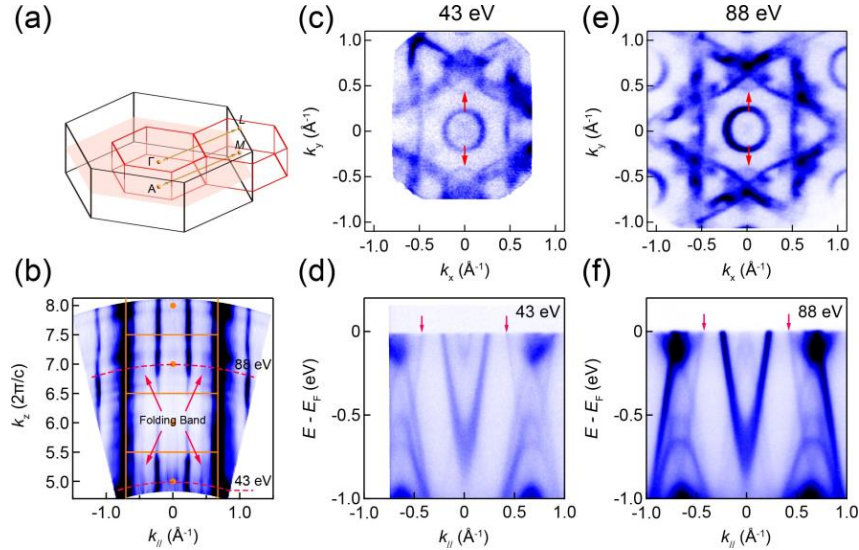

Figure S11: (a) Illustration of pristine three-dimensional Brillouin zone (dark frame) and 2×2×2 Brillouin zones (red frames), the scattering vectors are marked with double-headed yellow arrows. (b) The Fermi surface $k_z$-$k_{//}$ ARPES spectrum taken along $-M$ $\Gamma$ $M$ direction, the origin frames denote the edge of Brillouin zones, the red dash lines denote the 88 and 43 eV cuts, and the red arrows denote the folding bands. (c, e) Constant energy counter of ARPES mapping taken at $E_F$ with (c) 43 and (e) 88 photons. (d, f) The $E - k$ ARPES spectrum taken along $M - \Gamma - M$ direction at (d) 43 eV and (f) 88 eV

## 1.6 Temperature dependent spectra

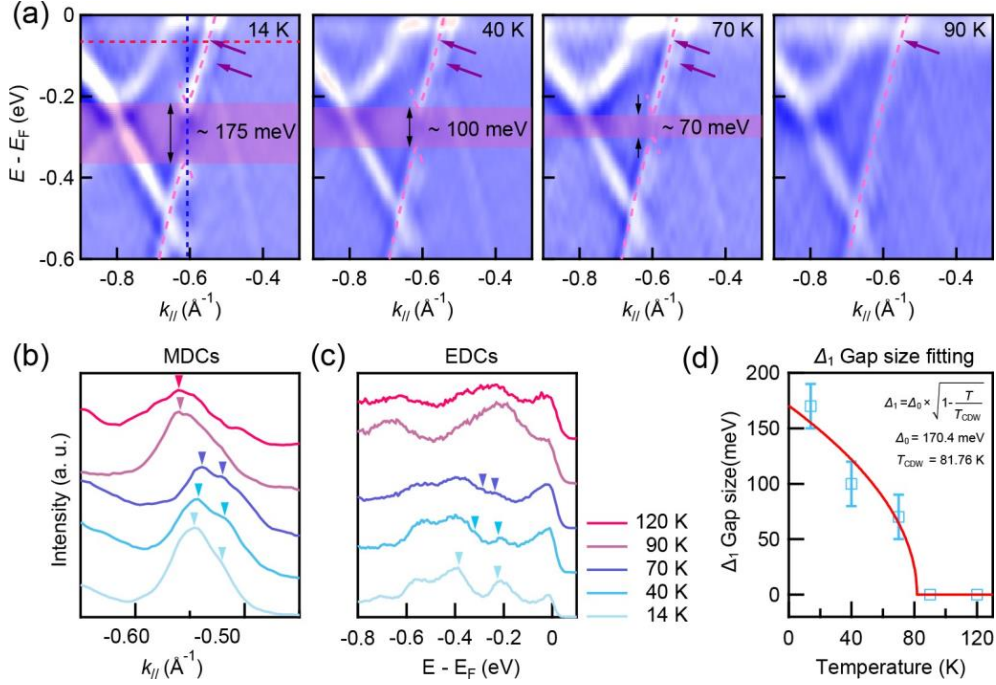

Figure S12: **Detail temperature dependent evolution of band splitting and folding band gap along *K-M* direction.** (a) Second derivation ARPES spectra of band dispersion along $K_1$-$\Gamma$ direction taken from 14 K to 90 K. Purple arrows denote the splitting $\alpha$ bands below $T_{CDW}$. Red shaded area denotes the CDW folding induced gap. (b) Momentum dispersion curves (MDCs) taken at $E_B$ = -0.05 eV and cross the $\alpha$ bands (marked by red dot line in (a)), the inverted triangular denotes the intensity peak in MDCs. (c) The energy distribution curves (EDCs) at the CDW gap $\Delta_1$ obtained from 14 to 120 K. (d) The temperature dependent CDW gap $\Delta_1$ fitting curve. The CDW gap $\Delta_1$ have the temperature dependent behavior with the BCS gap function, where is $\Delta_1 = \Delta_0 \times \sqrt{1 - \frac{T}{T_{CDW}}}$. $\Delta_0$ = 170 meV is the fitted gap size at 0 K and $T_{CDW}$ =81.8 K is the fitted CDW transition temperature.

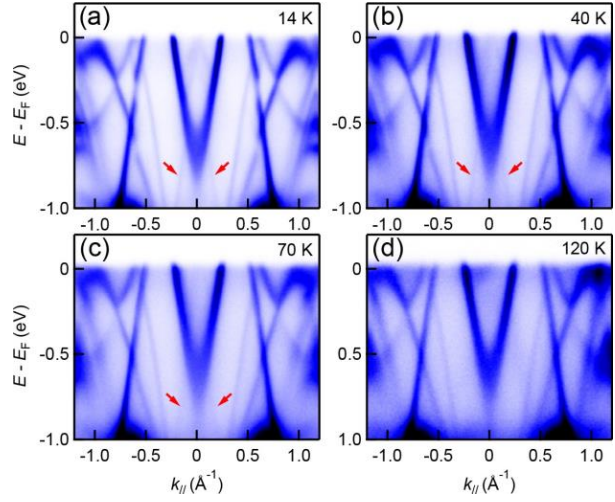

Figure S13: (a)-(d) ARPES spectra along $M_2$-$K_1$-$\Gamma$-$K_1$-$M_2$ direction taken at (a) 14 K, (b) 40 K, (c) 70 K, and (d) 120 K, respectively. Red arrows indicate the back-folded band near $\Gamma$ point.

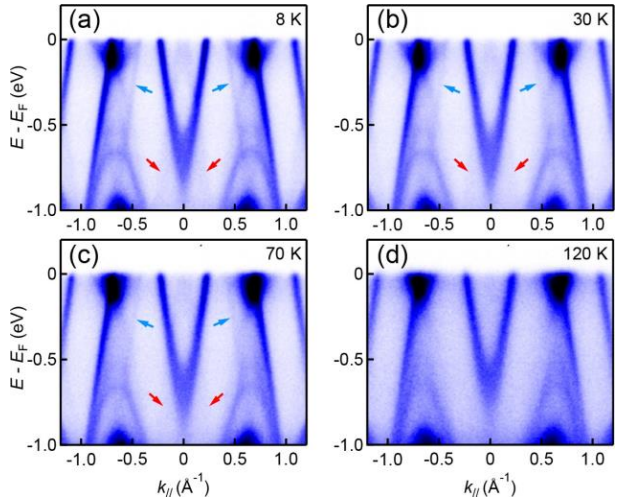

Figure S14: (a)-(d) ARPES spectra along $M_1$-$\Gamma$-$M_1$ direction taken at (a) 8 K, (b) 30 K, (c) 70 K, (d) 120 K, respectively. Red arrows indicate the back-folded band near $\Gamma$ point.

## 1.7 Constant energy map from high binding energy

$C_2$ symmetry with folding bands can only preserved in the region near to the Fermi level, as increasing the binding energy, the $C_6$ rotational symmetry recovered at around -1.0 eV. Given the comparison between the Fermi surface and the constant energy map at the higher binding energy of -1.0 eV, the change from $C_2$ to $C_6$ depending on the electron binding energy show the similar behavior with the previous STM results of $KV_3Sb_5$ and $CsV_3Sb_5$ below $T_{CDW}$.

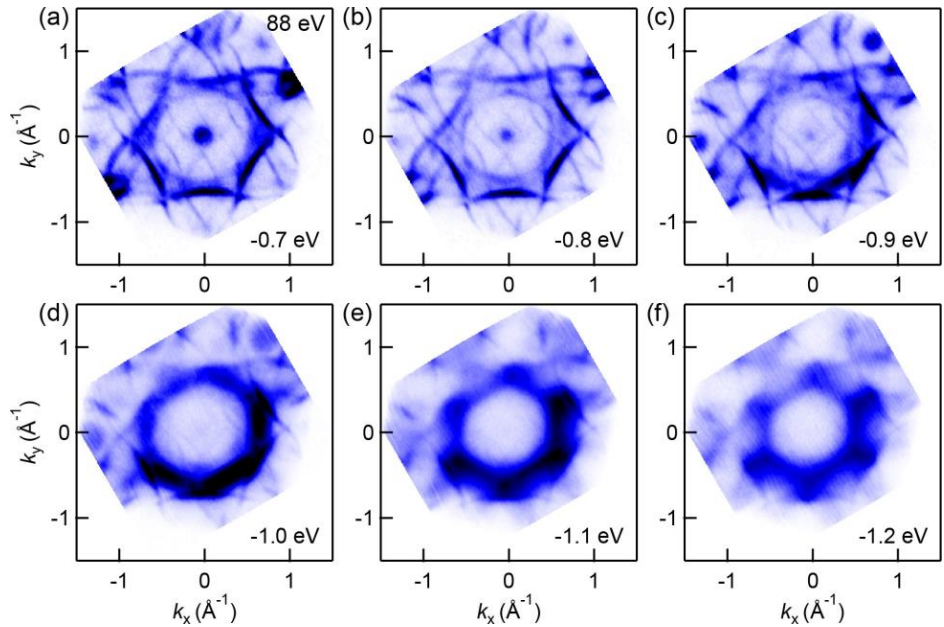

Figure S15: **Photoemission intensity map in the $k_x$-$k_y$ plane measured with 88 eV photon energy.** The constant energy map of (a) -0.7, (b) -0.8, (c) -0.9, (d) -1.0, (e) -1.1 and (f) -1.2 eV.

# 2. Calculation results

## 2.1 Comparison of 221 and 222 CDW state of KV$_3$Sb$_5$

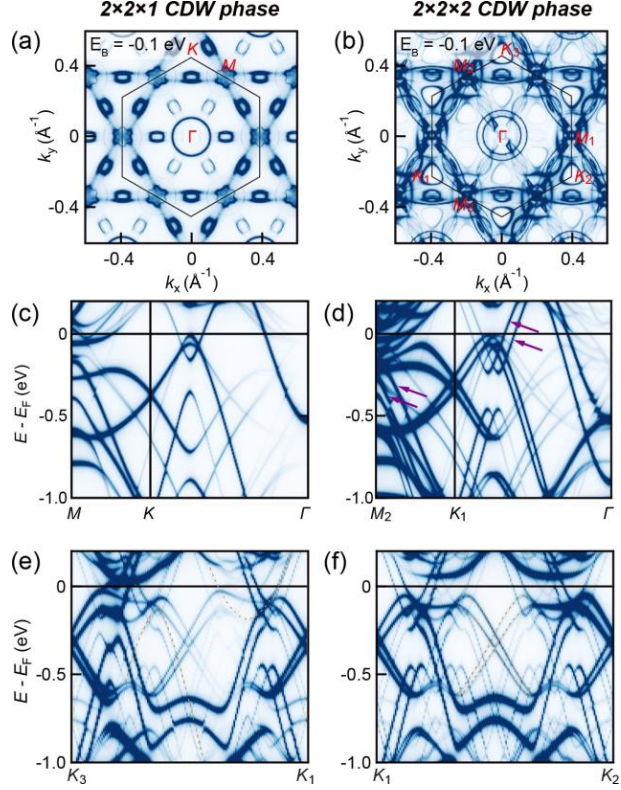

Figure S16: Difference between the band calculation of 2 × 2 × 1 and 2 × 2 × 2 CDW phase. (a), (b) Calculated constant energy surface mapping at $E_B$ = -0.1 eV for 2 × 2 × 1 and 2 × 2 × 2 CDW phase, respectively. (c), (d) Calculated band structure along *M-K-Γ* for 2 × 2 × 1 and *M$_2$-K$_1$-Γ* for 2 × 2 × 2 CDW phase, respectively. (e), (f) Calculated band structure along *K$_3$-K$_1$* and *K$_1$-K$_2$* directions. A dependence on the stacking configuration along the *c*-axis determine the rotational symmetry of the kagome lattice and the anisotropic dispersion of the electronic structure because of the quasi-1D electron scattering.

## 2.2 Continuum model for the anisotropic electronic structure in KV$_3$Sb$_5$

Here we construct a continuum model for the electronic nematic state in the 2 × 2 × 2 charge density wave (CDW) phase of KV$_3$Sb$_5$. To capture the essence of the electronic nematicity and the unidirectional band folding effects, here we only consider the hole pockets centered at the $\pm M_i$ ($i$ = 1, 2, 3) points and the electron pocket at $\Gamma$ point of the pristine Brillouin zone (BZ). There are two kagome layers in each primitive cell of the 2 × 2 × 2 CDW phase, and the two layers are mutually shifted by one of the primitive lattice vectors of the pristine structure (half of the CDW lattice vectors), denoted as $\mathbf{a}_i$ ($i$ = 1, 2, 3). Then, we construct the continuum model in the basis of the Bloch functions contributed by the two layers, and centered at the valleys $\Gamma$, and $\pm M_i$ ($i$ = 1, 2, 3) of the pristine BZ, which are denoted as $\psi_{n\mathbf{k}}^{(l)}(\mathbf{r})$. Here $n$ is the valley index (pocket index), $\mathbf{k}$ is the wavevector around each valley, and $l$ = 1, 2 is the layer index. Consider that layer 2 is shifted with respect to layer 1 by a pristine lattice vector $\mathbf{a}$, then Bloch function of the second layer is related to that of the first layer via:

$$\psi_{n\mathbf{k}}^{(2)}(\mathbf{r}) = \psi_{n\mathbf{k}}^{(1)}(\mathbf{r} - \mathbf{a}) = e^{i(\mathbf{k}+\mathbf{K}_n)\cdot(\mathbf{r}-\mathbf{a})} u_{n\mathbf{k}}^{(1)}(\mathbf{r} - \mathbf{a}) = e^{-i(\mathbf{k}+\mathbf{K}_n)\cdot\mathbf{a}} \psi_{n\mathbf{k}}^{(1)}(\mathbf{r}) , \qquad (S1)$$

where $\mathbf{K}_n$ = $\Gamma$, $\pm M_i$ ($i$ = 1, 2, 3) is the wavevector for the $n$th valley. The last equality in Eq. (S1) follows from the fact that the basis functions are Bloch functions of a lattice with in-plane pristine lattice vectors, such that the periodic part of the basis Bloch function $u_{n\mathbf{k}}(\mathbf{r})^{(l)} = u_{n\mathbf{k}}^{(l)}(\mathbf{r} - \mathbf{a})$. From Eq. (S1), it follows that the intra-layer Hamiltonian matrix elements can be expressed as:

$$\langle \psi_{n\mathbf{k}}^{(l)} | H | \psi_{n'\mathbf{k}}^{(l)} \rangle = E_{n\mathbf{k}}^0 \delta_{nn'} + \sum_{\mathbf{Q}} V_{nn'}^{(l)}(\mathbf{Q}) \delta_{\mathbf{K}_{n'}+\mathbf{Q}, \mathbf{K}_n} , \quad l = 1, 2 , \qquad (S2)$$

where $\mathbf{Q}$ denotes the reciprocal vector of the 2 × 2 × 2 CDW superlattice. The first term on the right hand side of Eq. (S2) is just the eigenenergy of the $n$th valley at $\mathbf{k}$, and the second term denotes the "intervalley" scattering induced by the inverse star-of-David (ISD) type of distortions within the kagome plane. Although the ISD distortion would change the electronic structure in a nontrivial way, e.g., giving rise to new van Hove singularities below the Fermi level, the threefold rotational symmetry is still preserved even in the presence of the ISD distortion. Therefore, the scatterings induced by the in-plane ISD distortions do not contribute to electronic nematicity observed in experiments, hence can be neglected in the spirit of leading-order approximation:

$$\langle \psi_{n\mathbf{k}}^{(l)}|H|\psi_{n'\mathbf{k}}^{(l)}\rangle \approx E_{n\mathbf{k}}^0 \delta_{nn'} \tag{S3}$$

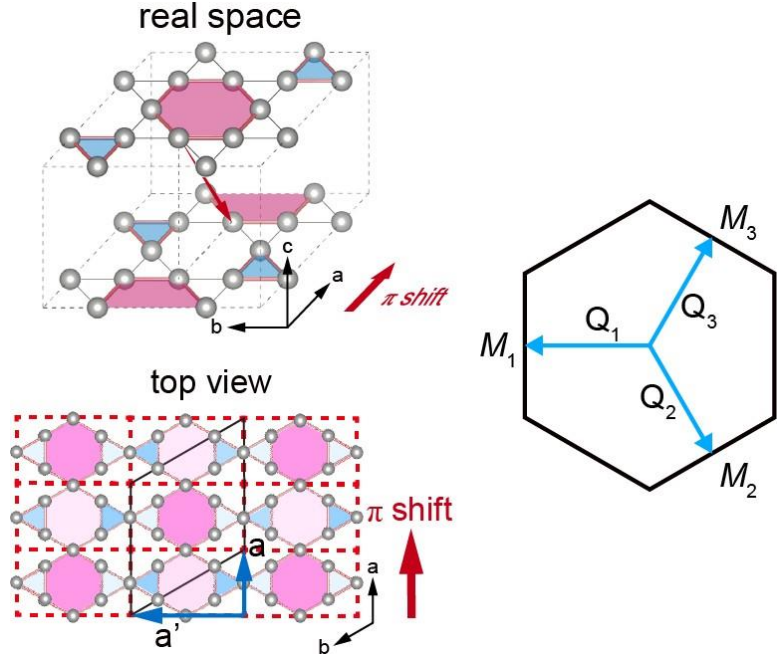

Figure S17: Schematic illustration of the lattice structures and the Brillouin zone.

Specifically, for the electron pocket at $\Gamma$, we take $E_{\mathbf{k}\Gamma}^0 = \hbar^2 k^2/(2m_\Gamma) + E_\Gamma$; and for the hole pockets at $\pm M_i$ ($i$ = 1, 2, 3), we take $E_{\mathbf{k}M} = \hbar^2 k^2/(2m_M) + E_M$. Next, we need to consider the interlayer Hamiltonian matrix element, i.e, the interlayer coupling term, some straightforward algebra shows that

$$\langle \psi_{n\mathbf{k}}^{(1)}|H|\psi_{n'\mathbf{k}}^{(2)}\rangle = \sum_{\mathbf{Q}} V_{nn'\mathbf{k}}^{\text{inter}}(\mathbf{Q}) e^{-i(\mathbf{k}+\mathbf{K}_n-\mathbf{Q})\cdot \mathbf{a}} \delta_{\mathbf{K}_{n'}+\mathbf{Q},\mathbf{K}_n}. \tag{S4}$$

Again, $\mathbf{Q}$ is the reciprocal vector of the 2 × 2 × 2 superlattice, and

$$V_{nn'\mathbf{k}}^{\text{inter}}(\mathbf{Q}) = H(\mathbf{Q}) \langle u_{n\mathbf{k}+\mathbf{K}_n}^{(1)}|u_{n'\mathbf{k}+\mathbf{K}_{n'}-\mathbf{Q}}^{(1)}\rangle \tag{S5}$$

denotes the interlayer scattering amplitude between valley $n$ and $n'$ with scattering wavevector $\mathbf{Q}$, where $H(\mathbf{Q}) = (1/V) \int_V d\mathbf{r}\, e^{-i\mathbf{Q}\cdot\mathbf{r}} H(\mathbf{r})$ is the Fourier transform of the real-space Hamiltonian $H(\mathbf{r})$.

Now we first consider the interlayer, and intravalley scattering, with $n = n'$ in Eq. (S5). If we only keep the dominant $\mathbf{Q} = 0$ component, we have

$$\langle \psi_{n\mathbf{k}}^{(1)} | H | \psi_{n\mathbf{k}}^{(2)} \rangle \approx H(0) e^{-i(\mathbf{k}+\mathbf{K}_n)\cdot \mathbf{a}}. \tag{S6}$$

For the interlayer, and intervalley scattering, it follows from Eq. (S5) that the $\mathbf{Q} = 0$ term vanishes, thus

$$\langle \psi_{n\mathbf{k}}^{(1)} | H | \psi_{n'\mathbf{k}}^{(2)} \rangle = \sum_{|\mathbf{Q}|\neq 0} V_{nn'\mathbf{k}}^{\text{inter}}(\mathbf{Q}) e^{-i(\mathbf{k}+\mathbf{K}_n - \mathbf{Q})\cdot \mathbf{a}} \delta_{\mathbf{K}_{n'}+\mathbf{Q}, \mathbf{K}_n}, \quad n \neq n' \tag{S7}$$

Without loss of generality, we consider that the second layer is shifted along the pristine lattice vector $\mathbf{a} = [a/2, \sqrt{3}a/2, 0]$, where $a$ is the lattice constant of the pristine structure. The lattice structure of the $2 \times 2 \times 2$ CDW phase is schematically shown in Figure S17. Let us define three in-plane primitive reciprocal vectors of the $2 \times 2 \times 2$ supercell, $\mathbf{Q}_1 = [-2\pi/\sqrt{3}a, 0, 0]$, $\mathbf{Q}_2 = [\pi/\sqrt{3}a, -\pi/a, 0]$, $\mathbf{Q}_3 = [\pi/\sqrt{3}a, \pi/a, 0]$, as shown in Figure S17. If one closely inspects the lattice structure of Figure S17, one finds that the lattice potential would have a strong modulation along the $\mathbf{a}$ direction with the modulation length of $|\mathbf{a}| = a$, which indicates the lattice potential would have relatively strong Fourier component at the wavevector $\mathbf{Q}_3 - \mathbf{Q}_2$ that is dual to $\mathbf{a}$, i.e., $\mathbf{a} \cdot (\mathbf{Q}_3 - \mathbf{Q}_2) = 2\pi$. Similarly, by inspection, the lattice potential would also have strong modulation along the direction perpendicular to $\mathbf{a}$, denoted as $\mathbf{a}'$, with a modulation length of $\sqrt{3}a$, which is dual to the reciprocal vector $\mathbf{Q}_1$ ($\mathbf{Q}_1 \cdot \mathbf{a}' = 2\pi$). This indicates that the Fourier component of the lattice potential at $\mathbf{Q}_1$ would be stronger than those of $\mathbf{Q}_2$ and $\mathbf{Q}_3$.

If we make a truncation to the summation over $\mathbf{Q}$ in Eq. (S7) such that all higher order Fourier components with $|\mathbf{Q}| > 2\pi/\sqrt{3}a$ are neglected, then Eq. (S7) only involves the scattering between the $\Gamma$ valley and the $M$ valleys, but with *anisotropic scattering amplitudes* for reasons explained above. Therefore, combining Eq. (S6), we have the following matrix elements for the interlayer couplings:

$$V^{\text{inter}}_{n;n,\mathbf{k}}(\mathbf{Q}=\mathbf{0}) = H(0) = V_0$$

$$V^{\text{inter}}_{\Gamma \pm M_1, \mathbf{k}}(\pm \mathbf{Q}_2) = V^{\text{inter}}_{\Gamma \pm M, \mathbf{k}}(\pm \mathbf{Q}_3) = V_2$$

$$V^{\text{inter}}_{\Gamma \pm M_2, \mathbf{k}}(\pm \mathbf{Q}_1) = V_1$$

$$V^{\text{inter}}_{nn,\mathbf{k}}(\mathbf{Q}) = 0, \quad \text{if } |\mathbf{Q}| > 2\pi/\sqrt{3}a. \tag{S8}$$

where $V_2 > V_1$ due to the stronger lattice potential modulation as discussed above. We set $V_1 = 0.12$ eV, and $V_0 = V_2 = 0.04$ eV. The effective mass for the electron (hole) pocket around the $\Gamma$ ($M$) point is $0.1674 m_0$ ($-0.0688 m_0$), where $m_0$ is the bare electron mass. The band minimum (maximum) energy for the $\Gamma$ ($M$) pocket is $E_\Gamma = -0.7565$ eV ($E_M = -0.0345$ eV). With Eq. (S3)-(S8), the continuum model has been well constructed. We then diagonalize the continuum Hamiltonian, and plot the unfolded spectral functions along the $-M_2 - \Gamma - M_2$, $-M_1 - \Gamma - M_1$, $-M_3 - \Gamma - M_3$ directions in Figure S18a, b, and c, respectively.

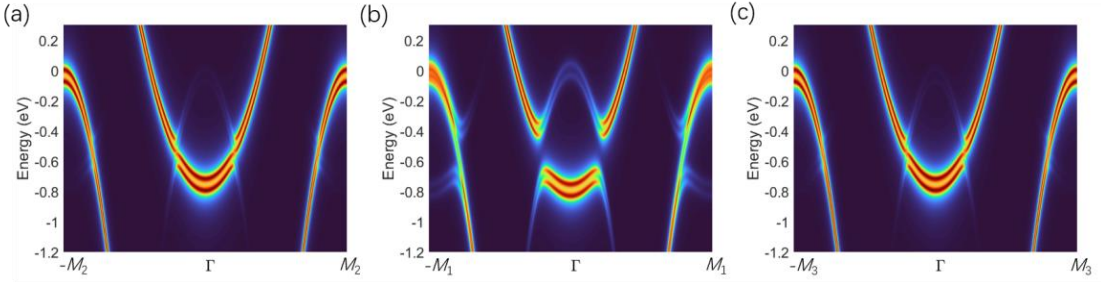

Figure S18: Unfolded spectral functions along the (a) $-M_1 - \Gamma - M_1$, (b) $-M_2 - \Gamma - M_2$, (c) $-M_3 - \Gamma - M_3$ directions. We set $V_2 = 0.12$ eV, and $V_1 = V_0 = 0.04$ eV.

## 3. Transport measurements

The single crystal was cut into long strips along the *ab* plane. The magneto-transport measurements were carried out by using a standard six-wire method in a physical property measurement system (PPMS, Quantum Design). During measurements, the magnetic field *B* was perpendicular to the *ab* plane and the electrical current *I* was along the *ab* plane.

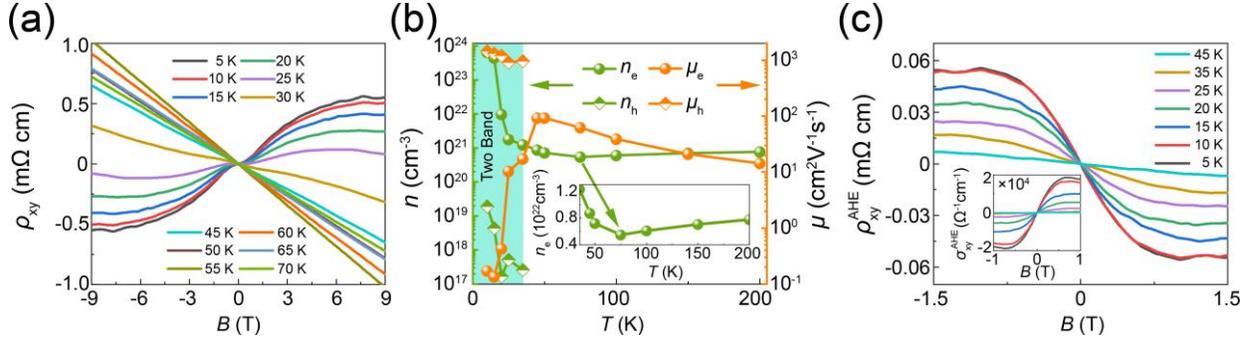

Figure S19: (a) Hall resistivity of $KV_3Sb_5$ with in-plane current and out-of-plane magnetic field. (b) Temperature dependent measurement of electron carrier concentration and mobility of $KV_3Sb_5$ (The data was fitted with two band model below 50 K). Inset: the enlarged curves around CDW transition temperature. (c) Extracted $\rho_{xy}^{AHE}$ and $\sigma_{xy}^{AHE}$ (inset) at various temperature.